# List-Mode PET Image Reconstruction Using Dykstra-Like Splitting

Kibo Ote, Fumio Hashimoto, Yuya Onishi, and Yasuomi Ouchi

*Abstract*— Convergence of the block iterative method in image reconstruction for positron emission tomography (PET) requires careful control of relaxation parameters, which is a challenging task. The automatic determination of relaxation parameters for list-mode reconstructions also remains challenging. Therefore, a different approach would be desirable. In this study, we propose a list-mode maximum likelihood Dykstra-like splitting PET reconstruction (LM-MLDS). LM-MLDS converges the list-mode block iterative method by adding the distance from an initial image as a penalty term into an objective function. LM-MLDS takes a two-step approach because its performance depends on the quality of the initial image. The first step uses a uniform image as the initial image, and then the second step uses a reconstructed image after one main iteration as the initial image. In a simulation study, LM-MLDS provided a better tradeoff curve between noise and contrast than the other methods. In a clinical study, LM-MLDS removed the false hotspots at the edge of the axial field of view and improved the image quality of slices covering the top of the head to the cerebellum. List-mode proximal splitting reconstruction is useful not only for optimizing nondifferential functions but also for converging block iterative methods without controlling relaxation parameters.

## I. INTRODUCTION

POSITRON emission tomography (PET) is an important tool for observing in vivo metabolism and elucidating disease mechanisms and brain functions [1]. In PET imaging, image reconstruction is essential to visualize the distribution of radiotracers within a living body using PET coincidence data. Because the number of counts per frame is limited when the dose is reduced to reduce exposure, or when measurements are repeated at short time intervals to track changes in the radiotracers over time, PET reconstruction is a battle against statistical noise. Incorporating the statistical and physical models into image reconstruction using iterative methods such as maximum likelihood expectation maximization (MLEM) [2] is crucial to providing a sufficient image for diagnosis within a limited dose and scan time. MLEM provides better image quality than analytical methods such as filtered backprojection (FBP); however, the computational amount ranges from tens to hundreds of times greater than FBP because it performs forward and backprojection operations at each iteration step. Block iterative methods such as ordered subset EM (OSEM) [3] have been developed to accelerate iterative methods. OSEM accelerates MLEM by dividing the sinogram into subsets depending on the projection angle, and updates the image using these subsets. Although the speed of the OSEM is enhanced by a factor equivalent to the number of subsets, the presence of statistical noise in the sinogram may hamper convergence to the maximum likelihood (ML) solution owing to the limit cycle phenomenon. To avoid the limit cycle phenomenon while facilitating faster convergence, a row-action ML algorithm (RAMLA) [4] and its subsequent method, the dynamic RAMLA (DRAMA) [5], have been developed. Both RAMLA and DRAMA maximize the number of subsets equal to the number of angles and introduce a relaxation parameter that reduces the magnitude of image updates as the iterations progress. In particular, DRAMA dynamically controls the relaxation parameter depending on the number of sub-iterations [5]. Although RAMLA and DRAMA are provably convergent [6], the optimal scheduling of the relaxation parameter for faster convergence remains a challenge.

In addition to the challenges of noise and computational time, PET reconstruction is a complicated task that increases the number of lines of response (LOR). With advances in hardware technologies, such as 3D-PET, time-of-flight (TOF), depth-of-interaction (DOI), and total-body PET, the number of LOR has increased exponentially [7], [8]. Storage of raw data without compression results in extremely large sinograms. Therefore, list-mode data acquisition and reconstruction have been introduced to leverage raw data without a sinogram. List-mode data acquisition encodes the various attributes of the detected coincidence event into bit-code and saves it into list data in the detected order. List-mode reconstruction directly generates an image from the list data without accumulating events in the sinograms [9], [10]. These list-mode technologies enable the maximization of the performance of state-of-the-art PET scanners without compromise. Therefore, a block iterative method for list-mode PET reconstruction is essential for advanced PET scanners. The list-mode and sinogram-based block iterative reconstruction methods use different data spaces when dividing the emission data into subsets. The sinogram-

K. Ote, F. Hashimoto, and Y. Onishi are with Central Research Laboratory, Hamamatsu Photonics K. K., 5000 Hirakuchi, Hamana-ku, Hamamatsu 434-8601, Japan. (e-mail: kibou@crl.hpk.co.jp, fumio.hashimoto@crl.hpk.co.jp, yuya.onishi@hpk.co.jp).

Y. Ouchi is with Department of Biofunctional Imaging, Preeminent Medical Photonics Education & Research Center, Hamamatsu University School of Medicine, 1-20-1 Handayama, Chūō-ku, Hamamatsu 431-3192, Japan. (e-mail: ouchi@hama-med.ac.jp).
.

based block iterative method divides the sinogram into subsets based on the angle of projection, whereas the list-mode block iterative method divides the list data into subsets based on the order of event detection. Although the optimal relaxation parameters for sinogram-based DRAMA can be determined automatically based on the geometric correlation between the projection angles [5], there is no method to automatically determine the optimal relaxation parameters for list-mode reconstruction. In practice, the optimal relaxation parameters in list-mode reconstruction are determined experimentally based on image quality evaluation. However, there is no guarantee that these parameters are optimal for unknown data.

In recent years, the proximal splitting method has been applied to various fields of signal and image processing [11]. The proximal splitting method is primarily used for the following purposes. First, it is used to solve problems that are difficult to optimize using a gradient descent method such as total variation (TV) minimization [12]. Second, it is used to divide an objective function consisting of multiple loss functions into simpler subproblems [13]–[15]. In the field of PET image reconstruction, the proximal splitting method has been used to integrate deep learning and PET reconstruction [16], [17], and to construct a block iterative method for faster PET reconstruction incorporating total variation minimization [18], [19]. In addition, it has been proposed to use the proximal splitting method as a unified approach to derive various block iterative methods for computed tomography (CT) and PET reconstruction [20], [21].

In this study, we propose a novel block iterative method for list-mode maximum likelihood PET reconstruction using Dykstra-like splitting [11], [14] (LM-MLDS). Dykstra-like splitting constructs a block iterative method with a proximity operator by imposing a penalty on the distance from the initial reference image. The proximity operator aims to minimize the loss function as much as possible near a given point. Dykstra-like splitting likely converges without controlling the relaxation parameter, because the image update automatically decreases as the distance from the reference image increases. Therefore, we anticipate that a proximal splitting-based block iterative method for list-mode PET reconstruction will suppress the limit cycle phenomenon and yield improved convergence compared to list-mode OSEM (LM-OSEM). We used this framework to construct a new list-mode block iterative method. We evaluated the proposed LM-MLDS using simulations and clinical data from a brain PET scanner with four-layer DOI detectors [22]. LM-MLDS showed better tradeoff curves between noise and contrast than LM-OSEM and LM-DRAMA and showed different noise properties induced by the proximity operator. These results suggest that proximal splitting methods, such as Dykstra-like splitting, are useful not only for optimizing nondifferentiable functions, but also as a framework for constructing block iterative methods for list-mode PET reconstruction.

## II. BACKGROUND

In this section, we briefly introduce the list-mode PET reconstruction. List data can be expressed as a sequence of LOR indices that detect the coincidence event [23].

$$D = \{i(t) | t = 0, 1, \cdots, T-1\}, \quad (1)$$

where $t$ is the detected order of a coincidence event, $T$ is the total number of coincidence events, and $i(t)$ is an index of LOR that detects the $t^{\text{th}}$ event.

The list-mode log-likelihood function to be maximized in image reconstruction is expressed as [24]

$$L(D|x) = \sum_t \log\left\{\sum_j a_{i(t)j} x_j + \bar{s}_{i(t)}\right\} - \sum_{i,j} a_{ij} x_j, \quad (2)$$

where $x$ is the image of the subject, $\bar{s}$ is the scatter estimate, $j$ is the index of the voxel, $a_{ij}$ is an element of the system matrix expressing the contribution of $j^{\text{th}}$ voxel to $i^{\text{th}}$ LOR, and $a_{i(t)j}$ is the element of system matrix corresponding to the $t^{\text{th}}$ event.

List-mode MLEM (LM-MLEM) reconstructs an image from list data using the following recursive formula:

$$x_j^{(k+1)} = \frac{x_j^{(k)}}{\sum_i a_{ij}} \sum_t a_{i(t)j} \frac{1}{\sum_{j'} a_{i(t)j'} x_{j'}^{(k)} + \bar{s}_{i(t)}}, \quad (3)$$

where $k$ is an iteration number.

For block optimization, events are partitioned into $M$ subsets as follows [25]:

$$T_0 = \{t | t\%M = 0\}, T_1 = \{t | t\%M = 1\},$$
$$\cdots, T_{M-1} = \{t | t\%M = M-1\}. \quad (4)$$

We consider a block optimization problem as,

$$\min_{x \in \mathbb{R}_+^J} \sum_q -L_q(D|x), \quad (5)$$

$$L_q(D|x) = \sum_{t \in T_q} \log\left\{\sum_j a_{i(t)j} x_j + \bar{s}_{i(t)}\right\} - \frac{\sum_{i,j} a_{ij} x_j}{M}, \quad (6)$$

where $L_q(D|x)$ is the list-mode log-likelihood function for the $q^{\text{th}}$ subset, $\mathbb{R}_+^J$ is a set of $J$-dimensional vecotrs of postive real numbers, and $J$ is a number of voxels.

LM-OSEM solves the problem in Eq. (5) using a recursive formula [9]:

$$x_j^{(k,q+1)} = \frac{x_j^{(k,q)}}{\omega_j} \sum_{t \in T_q} a_{i(t)j} \frac{1}{\sum_{j'} a_{i(t)j'} x_{j'}^{(k)} + \bar{s}_{i(t)}}, \quad (7)$$

$$\omega_j = \frac{1}{M} \sum_i a_{ij}, \quad (8)$$

where $k$ and $q$ are the main- and sub-iteration numbers, and $\omega_j$ is a sensitivity image [26].

Next, LM-DRAMA solves the optimization problem in Eq. (5) using the following recursive formula:

$$x_j^{(k,q+1)} = x_j^{(k,q)} + \lambda^{(k,q)} x_j^{(k,q)} \left( \frac{1}{\omega_j} \sum_{t \in T_q} a_{i(t)j} \frac{1}{\sum_{j'} a_{i(t)j'} x_{j'}^{(k,q)} + \bar{s}_{i(t)}} - 1 \right), \quad (9)$$

$$\lambda^{(k,q)} = \frac{\beta}{\beta + q + \gamma k M}, \quad (10)$$

where $\lambda^{(k,q)}$ is a subset-dependent relaxation parameter. $\beta$ and $\gamma$ are the positive constants controlling the relaxation parameter.

### III. PROPOSED METHOD

In this section, we introduce the proposed LM-MLDS method in three subsections: Dykstra-like splitting, proximity operator, and overall algorithm.

#### A. Dykstra-like splitting

List-mode PET reconstruction using Dykstra-like splitting considers the following optimization problem:

$$\min_{x \in \mathbb{R}_+^J} \frac{1}{2\alpha} \|x - r\|^2 + \sum_q -L_q(D|x), \quad (11)$$

where $r$ is a reference image and $\alpha$ is a step size.

Following an algorithm for Dykstra-like splitting [11, 21], we solve the above optimization problem as

$$x^{(0,0)} = r, \quad y_0^{(0)} = y_1^{(0)} = \cdots = y_{M-1}^{(0)} = 0, \quad (12)$$

$$x^{(k,q+1)} = \text{prox}_{-\alpha L_q}\left(x^{(k,q)} + y_q^{(k)}\right)$$
$$= \underset{x \in \mathbb{R}_+^J}{\text{argmin}} \left\{ -L_q(D|x) + \frac{1}{2\alpha} \left\| x - \left(x^{(k,q)} + y_q^{(k)}\right) \right\|^2 \right\}, \quad (13)$$

$$y_q^{(k+1)} = x^{(k,q)} + y_q^{(k)} - x^{(k,q+1)}, \quad (14)$$

where $\text{prox}_{-\alpha L_q}(x^{(k,q)} + y_q^{(k)})$ is a proximity operator minimizing $-L_q(D|x)$ in the proximity of $x^{(k,q)} + y_q^{(k)}$, and $y_q$ is a dual variable for the $q^{\text{th}}$ subset which maintains the difference between $x^{(k,q)}$ and $x^{(k,q+1)}$. In Dykstra-like splitting, the reference image $r$ becomes an initial image $x^{(0,0)}$ as shown in Eq. (12). We set the reference image $r$ to a uniform image with a voxel value of one, the same as that in conventional ML methods.

#### B. Proximity operator

The specific calculation of the proximity operator in Eq. (13) can be derived using the optimization transfer method [27], [28]. To update each voxel independently, we construct the surrogate function of $L_q$ as

$$g_q\left(x \middle| x_{\text{EM}}^{(k,q+1)}\right) = \sum_j \omega_j \left( x_{j,\text{EM}}^{(k,q+1)} \log x_j - x_j \right), \quad (15)$$

where $g_q$ is a surrogate function of $L_q$ and $x_{\text{EM}}$ is an updated image using the LM-OSEM's recursive formula of Eq. (7).

The surrogate function $g_q$ satisfies the following two conditions.

$$g_q\left(x \middle| x_{\text{EM}}^{(k,q+1)}\right) - g_q\left(x_{\text{EM}}^{(k,q+1)} \middle| x_{\text{EM}}^{(k,q+1)}\right)$$
$$\leq L_q(D|x) - L_q\left(D \middle| x_{\text{EM}}^{(k,q+1)}\right), \quad (16)$$

$$\nabla g_q\left(x_{\text{EM}}^{(k,q+1)} \middle| x_{\text{EM}}^{(k,q+1)}\right) = \nabla L_q\left(D \middle| x_{\text{EM}}^{(k,q+1)},\right). \quad (17)$$

Hence, maximizing the surrogate function $g_q$ maximizes the original objective function $L_q$. Replacing $L_q$ in Eq. (13) with $g_q$, we obtain the surrogate objective function, which can be optimized voxel-by-voxel as follows:

$$P\left(x_j \middle| x_{\text{EM}}^{(k,q+1)}\right) = -\omega_j \left( x_{j,\text{EM}}^{(k,q+1)} \log x_j - x_j \right)$$
$$+ \frac{1}{2\alpha} \left\{ x_j - \left(x_j^{(k,q)} + y_{q,j}^{(k)}\right) \right\}^2, \quad (18)$$

By setting the derivative of Eq. (18) to zero, the following solution is obtained:

$$x_j^{(k,q+1)} = \frac{c_j^{(k,q)} + \sqrt{c_j^{(k,q)^2} + 4 x_{j,\text{EM}}^{(k,q+1)} \alpha \omega_j}}{2}, \quad (19)$$

$$c_j^{(k,q)} = x_j^{(k,q)} + y_{q,j}^{(k)} - \alpha \omega_j. \quad (20)$$

The proximity operator in Eq. (13) consists of one sub-iteration of LM-OSEM and regularization using Eq. (19).

#### C. Overall algorithm

The LM-MLDS algorithm is presented in **Algorithm 1**. LM-MLDS iterates LM-OSEM, the regularization using Eq. (19), and dual variable update. We begin updating the dual variables after one main iteration, as shown in lines 5–8 of **Algorithm 1**. This is equivalent to replacing the reference image with a reconstructed image after one main iteration, and resetting the dual variables to zero. We adopted this two-step approach because the performance of LM-MLDS depends on the reference image. In addition, we randomly permute the access order of each subset in every main iteration. The computational cost of LM-MLDS is similar to LM-OSEM because the comoutational cost of regularization is almost negilable, but it requires an additional memory to store the dual variables.

**Algorithm 1** Algorithm of LM-MLDS

Input: Iteration number $N$, Number of subset $M$, Step size $\alpha$, Sensitivity image $\omega$, Scatter estimate $\bar{s}$

Initialize: $x^{(0,0)}$ (uniform), $y_0^{(0)} = \cdots = y_{M-1}^{(0)} = 0$

1: **for** $k = 0$ **to** $N - 1$ **do**
2:    order = permutation$(0, M - 1)$
3:    **for** $l = 0$ **to** $M - 1$ **do**
4:      $q = \text{order}(l)$
5:      $x_{j,\text{EM}}^{(k,l+1)} = \frac{x_j^{(k,l)}}{\omega_j} \sum_{t \in T_q} a_{i(t)j} \frac{1}{\sum_{j'} a_{i(t)j'} x_{j'}^{(k,l)} + \bar{s}_{i(t)}}$
6:      $c_j^{(k,l)} = x_j^{(k,l)} + y_{q,j}^{(k)} - \alpha \omega_j$
       $x_j^{(k,l+1)} = \frac{1}{2} \left\{ c_j^{(k,l)} + \sqrt{{c_j^{(k,l)}}^2 + 4 x_{j,\text{EM}}^{(k,l+1)} \alpha \omega_j} \right\}$
7:      **if** $k > 0$ **then**
8:        $y_q^{(k+1)} = x^{(k,l)} + y_q^{(k)} - x^{(k,l+1)}$
9:      **else**
10:      $y_q^{(k+1)} = y_q^{(k)}$
11:    **end for**
12: **end for**
13: **return** $x^{(N-1,M-1)}$

## IV. EXPERIMENTAL SETUP

We used a brain PET scanner with a four-layer DOI (Hamamatsu HITS-655000) [22] for both the simulation and clinical experiments. The scanner consists of 32 (radial) × 5 (axial) detector units that consist of 32 × 32 lutetium-yttrium oxyorthosilicate (LYSO) scintillator array with 1.2 mm pitch and 8 × 8 array of multipixel photon counters (MPPC). The thickness of each DOI layer was designed to be 3, 4, 5, and 8 mm from the side closest to the subject to the side farthest from the subject. The number of crystals was $655 \times 10^3$ and the corresponding number of LOR was $107 \times 10^9$.

For image reconstruction, we performed LM-MLEM for 200 iterations and LM-OSEM, LM-DRAMA, and LM-MLDS with 40 subsets for five main iterations for both simulation and clinical data. We set $\beta$=40 and $\gamma$=0.1 for LM-DRAMA, and $\alpha$=2 for LM-MLDS. The $\alpha$ was determined experimentally. The image and voxel sizes were 128 × 128 × 83 and 2.6 mm × 2.6 mm × 2.4 mm, respectively. For attenuation correction, simulation studies used phantom attenuation maps, whereas clinical studies estimated attenuation maps by segmenting nonattenuation-corrected images [29]. Other data corrections were performed using component-based normalization [30], single-scatter simulation [31], and delayed coincidence subtraction [32]. In addition, we used a 3D Gaussian function with one-voxel full-width-at-half-maximum (FWHM) as a shift-invariant image-space point spread function (PSF) in list-mode PET reconstruction [9].

### A. Simulation data

The simulation data were generated using an in-house Monte Carlo simulation code with scatter and attenuation. A segmented image of magnetic resonance imaging (MRI) was downloaded from BrainWeb [33] to create a digital brain phantom. We set the activities of 1:0.25:0.05 for gray matter (GM), white matter (WM), and cerebrospinal fluid (CSF), respectively, and embedded three tumor regions with activities of 1.5, 1.2, 1.1 and radii of 1.0, 1.2, 1.6 cm in the digital brain phantom. We set the attenuation coefficient of 0.0151 mm$^{-1}$ and 0.00958 mm$^{-1}$ for bone and the other tissues, respectively. The simulated list data had $1.52 \times 10^8$ events. A low-count version of the simulation data was created by thinning the list data to 1/20 count.

For a quantitative evaluation, we measured the peak signal-to-noise ratio (PSNR) and tumor uptake ratio (TR) as follows:

$$\text{PSNR} = 10 \log_{10} \frac{\left( \max_{j \in R_{\text{brain}}} x_{j,\text{pha}} \right)^2}{\frac{1}{N_{\text{brain}}} \sum_{j \in R_{\text{brain}}} (x_j - x_{j,\text{pha}})^2}, \quad (21)$$

$$\text{TR} = \frac{\sum_{j \in R_{\text{tumor}}} x_j}{\sum_{j \in R_{\text{tumor}}} x_{j,\text{pha}}}, \quad (22)$$

where $x_{\text{pha}}$ is the phantom image, $R_{\text{brain}}$ is the region of interest (ROI) of the whole brain, $N_{\text{brain}}$ is the number of voxels inside the $R_{\text{brain}}$, and $R_{\text{tumor}}$ is the ROI of the tumor region.

To compare the dependence of noise on the local activity level, we set the ROIs on the WM and GM regions based on the voxel value of the digital brain phantom and determined the coefficient of variation (COV) as

$$\text{COV}_{\text{ROI}} = \frac{\text{StdDev}_{\text{ROI}}}{\text{Mean}_{\text{ROI}}} \times 100\%, \quad (23)$$

where $\text{StdDev}_{\text{ROI}}$ and $\text{Mean}_{\text{ROI}}$ are the standard deviation and mean values of the ROI, respectively.

### B. Clinical data

Clinical data were obtained from Hamamatsu University School of Medicine using a HITS-655000 scanner. The Ethics Committee of Hamamatsu University School of Medicine approved the study, and written informed consent was obtained from all participants prior to enrollment. A healthy volunteer was scanned 62 min after injection of 5 MBq/kg of $^{11}$C-MeQAA which is a tracer for α7 nicotinic acetylcholine receptors (nAChR) highly existing on the thalamus and striatum [34]. We employed a 42–62 min frame to obtain a clear accumulation of $^{11}$C-MeQAA in the thalamus. The list data of the 42–62 min frame had $4.7 \times 10^7$ events. A low-count version of the clinical data was created by thinning the list data to 1/10 count.

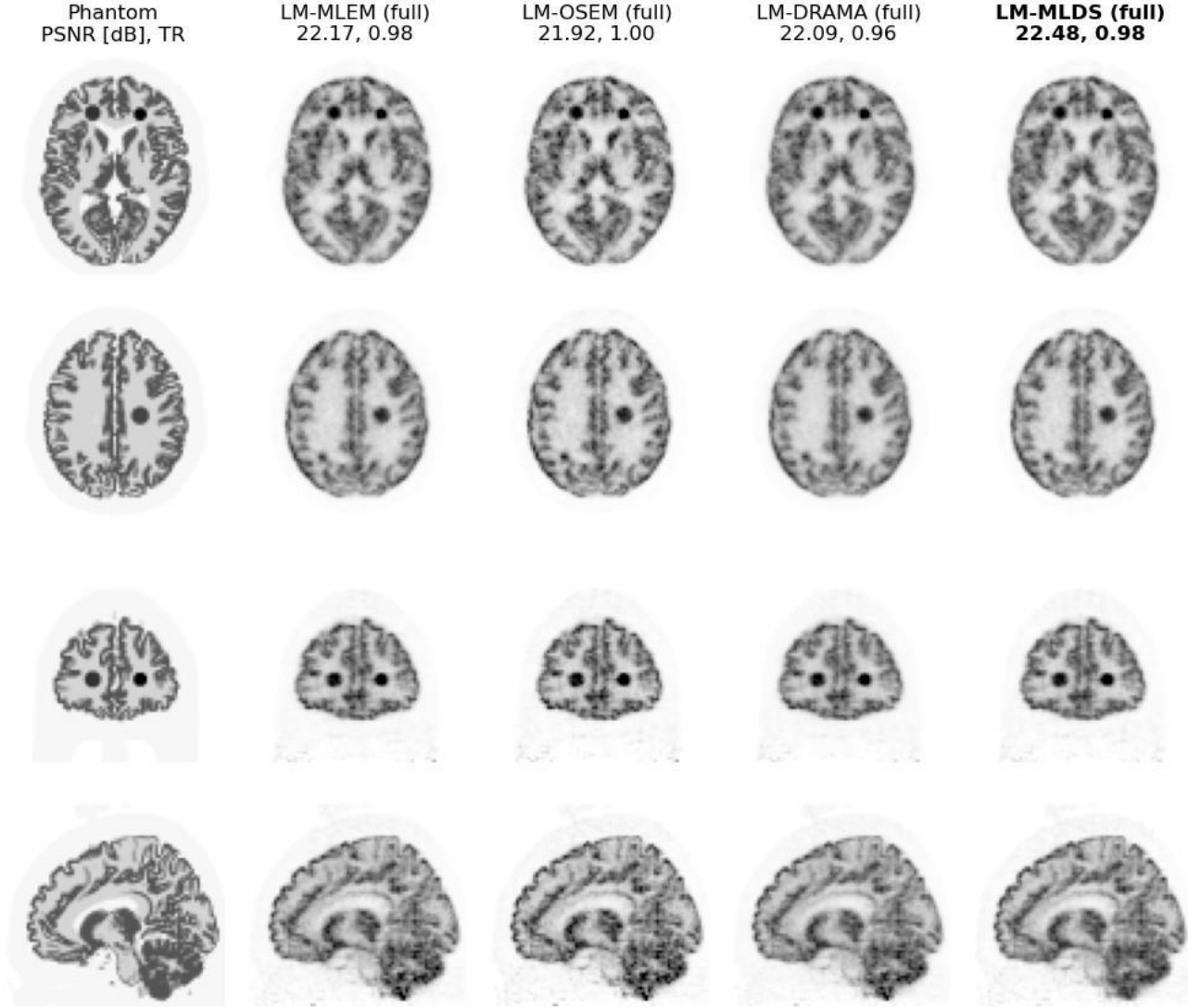

**Fig. 1.** Images of full count simulation data with three tumor regions reconstructed using the proposed method and the other methods. From left to right, phantom, LM-MLEM, LM-OSEM, LM-DRAMA, and LM-MDLS. Each image is tagged with PSNR and tumor uptake ratio. We show the images of LM-MLEM at 60 iterations, LM-OSEM, LM-DRAMA, and LM-MLDS at 2 main iteration.

As an indicator of signal, we evaluated uptake in the thalamus as a mean voxel value in the thalamus ROI. As an indicator of noise, we evaluated the normalized standard deviation (NSTD) between the ROI values of the WM as:

$$\text{NSTD} = \frac{1}{\bar{b}} \sqrt{\frac{1}{K_b} \sum_{k=1}^{K_b} (b_k - \bar{b})^2}, \quad (24)$$

where $K_b$ is the number of ROIs in WM, $b_k$ is $k^{\text{th}}$ ROI value in WM, and $\bar{b}$ is the mean of the ROI values in WM as

$$b_k = \frac{1}{N_{b,k}} \sum_{j \in R_{b,k}} x_j, \quad (25)$$

$$\bar{b} = \frac{1}{K_b} \sum_{k=1}^{K_b} b_k, \quad (26)$$

where $R_{b,k}$ is the $k^{\text{th}}$ ROI in WM. In this study, we set $K_b = 25$.

In addtion, we set the ROI on the cerebellum and determined the COV using Eq. (23).

## V. RESULTS

**Fig. 1** shows the reconstructed images of the simulation data with a full count. LM-OSEM provides a slightly noisier image than LM-MLEM. LM-DRAMA provides less-noisy and smoother images than LM-OSEM because of its relaxation parameters. LM-MLDS provides sharper images than LM-DRAMA, less noisy images than LM-OSEM, and achieves the highest PSNR.

**Fig. 2** shows the reconstructed images of the simulation data with a 1/20 count. LM-OSEM provides noisier images than LM-MLEM, and more noticeably than the full-count case. This was probably because the amplitude of the limit cycle phenomenon in the 1/20 count was larger than that in the full count. LM-DRAMA provides images with lower contrast than LM-OSEM, as indicated by the greyish WM regions. This is because the relaxation parameters converged slowly when the numbers of the subsets and iterations were the same. LM-MLDS provides less noise than LM-OSEM, higher contrast than LM-DRAMA, and achieves the highest PSNR. The PSNR are 17.85, 16.38, 18.16, 18.55 and the TR are 0.86, 0.90, 0.83, 0.88 for the LM-MLEM, LM-OSEM, LM-DRAMA, and LM-MLDS, respectively.

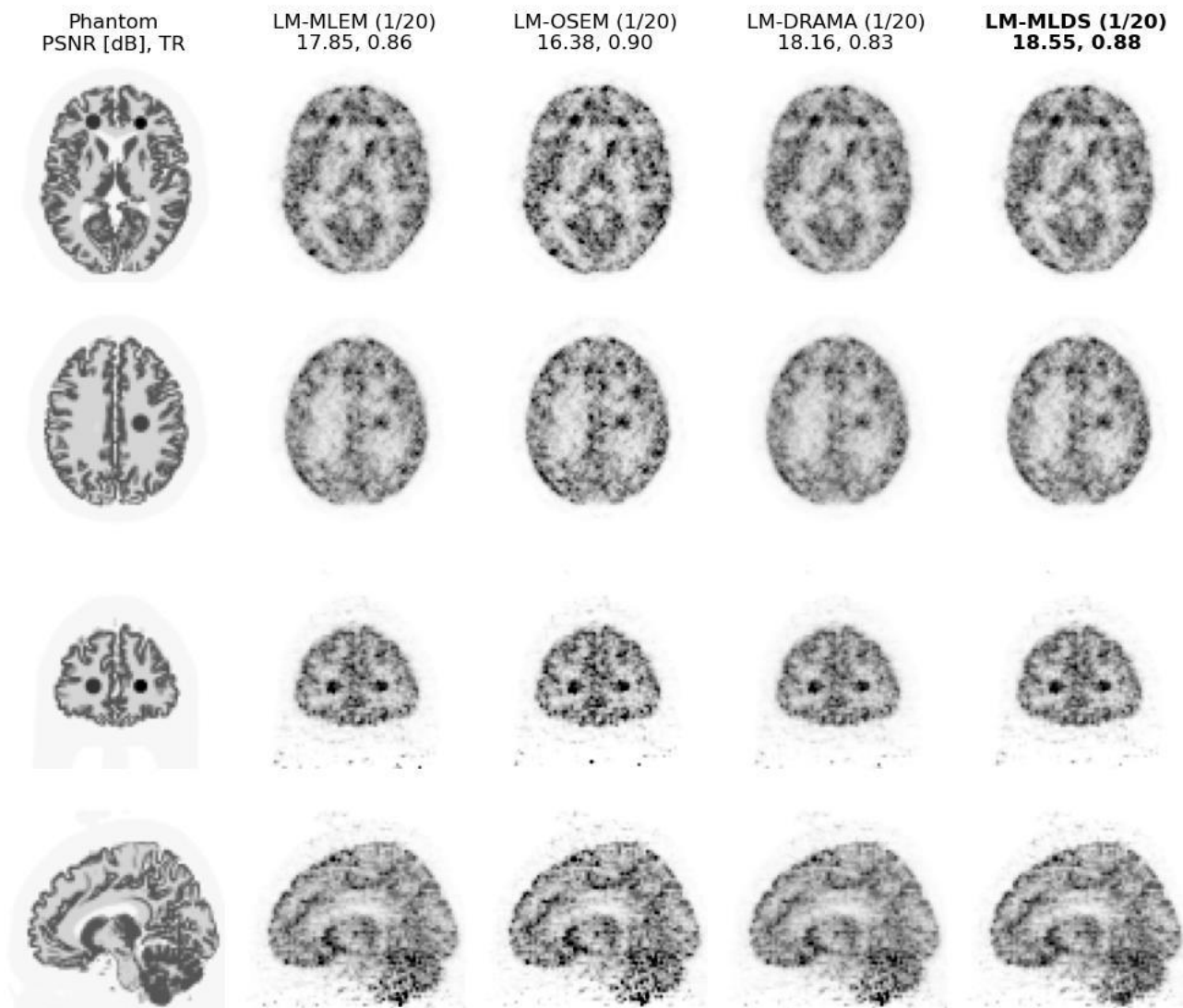

**Fig. 2.** Images of 1/20 count simulation data with three tumor regions reconstructed using the proposed method and the other methods. From left to right, phantom, LM-MLEM, LM-OSEM, LM-DRAMA, and LM-MLDS. Each image is tagged with PSNR and tumor uptake ratio. We show the images of LM-MLEM at 30 iterations, LM-OSEM, LM-DRAMA, and LM-MLDS at 1 main iteration.

**Fig. 3 left** shows the tradeoff curves between PSNR and TR after 40 updates in the simulation data with a full count. As a general trend, both PSNR and TR improve in early updates, but after some turning points, noise becomes dominant and the PSNR starts to decrease. In later iterations, TR slightly exceeds 1.0, except for LM-MLDS. The reason for this slight overshoot may be the ringing artifacts caused by PSF modeling [35]. The tradeoff curve of LM-MLEM is smooth, but that of LM-OSEM exhibits slight oscillations owing to the limit cycle phenomenon. LM-DRAMA reduces the amplitude of oscillation at later iterations but exhibits a wavy curve at early iterations. LM-MLDS exhibits a tradeoff curve with a higher PSNR to the right than the other methods. In addition, it suppresses TR overshoot and oscillations caused by the limit cycle phenomenon.

**Fig. 3 right** shows the tradeoff curves between PSNR and TR after 10 updates in simulation data with a 1/20 count. LM-MLEM provides a smooth tradeoff curve, but the PSNR decreases at an earlier iteration than in the full-count case. LM-OSEM provides greater oscillation of the tradeoff curve than in the full-count case. This may suggest that the degree of limit cycle depends on the number of counts per subset. LM-DRAMA suppresses the oscillation of the tradeoff curve slightly compared with LM-OSEM. LM-MLDS could not completely suppress the oscillation of the tradeoff curve, but gave a point closer to the top right of the graph compared with the other methods. The same hyperparameters of $\alpha$, $\beta$, and $\gamma$ for LM-MLDS and LM-DRAMA were used in both the full and 1/20 count cases.

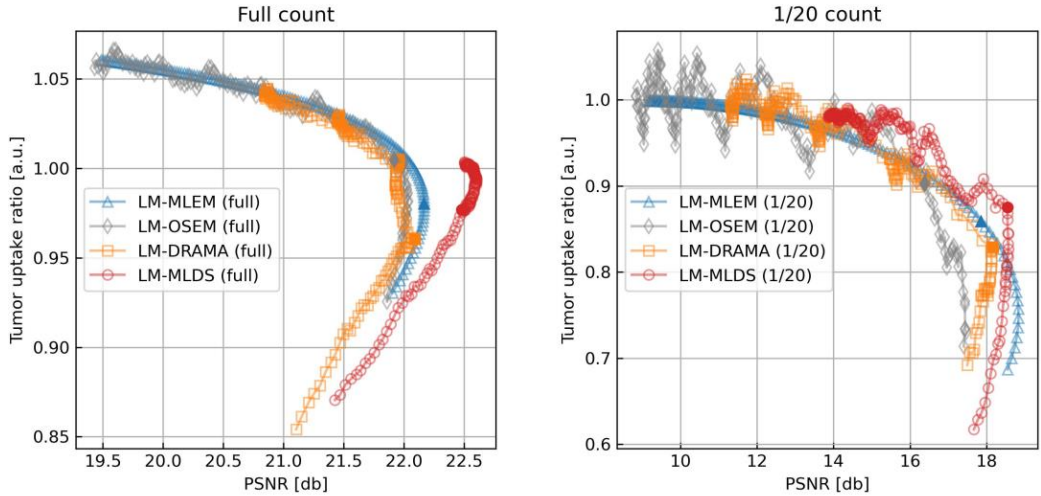

**Fig. 3.** Tradeoff curves between tumor uptake ratio and PSNR on simulation data in (left) full and (right) 1/20 count, respectively. Markers correspond to 40–200 and 10–200 updates for full and 1/20 count, respectively. Fill markers correspond to the images shown in Fig. 1 and Fig. 2.

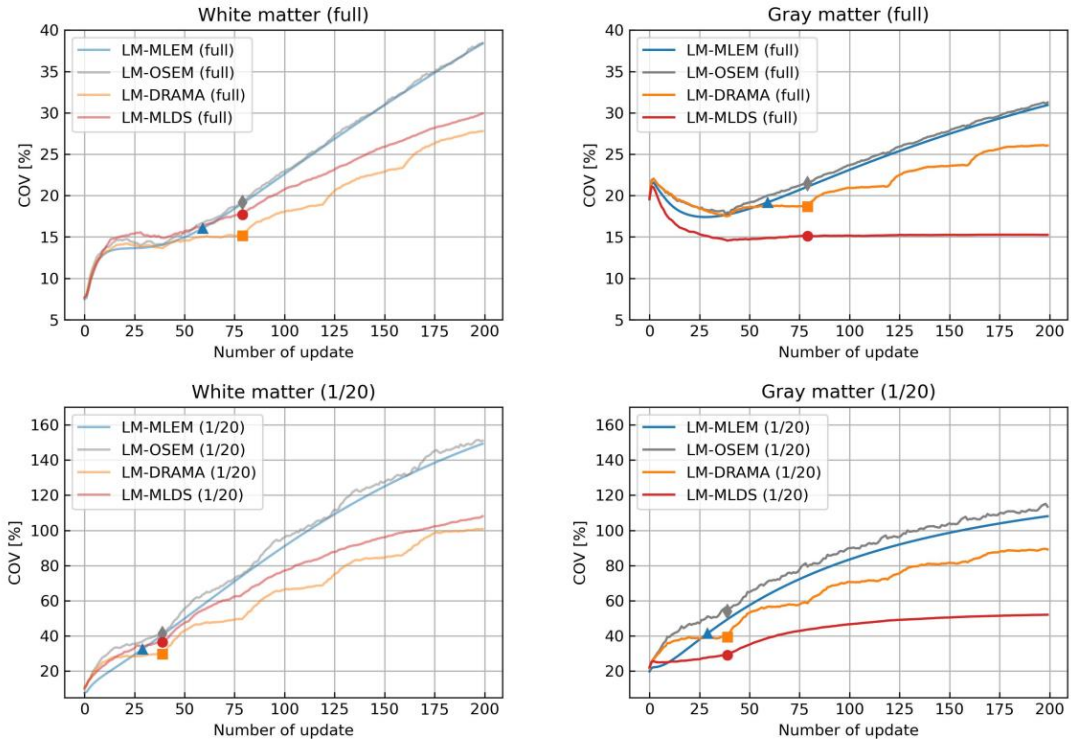

**Fig. 4.** COV curves on (left) white matter and (right) gray matter, respectively, for simulation data with (top) full count and (bottom) 1/10 count, respectively. Markers correspond to the images shown in Fig. 1 and Fig. 2.

**Fig. 4 top left** shows the $COV_{WM}$ curves in the simulation data with a full count. The $COV_{WM}$ curve for LM-OSEM is similar to that of LM-MLEM at later iterations. LM-DRAMA exhibits a lower and stepped $COV_{WM}$ curve in comparison to the other methods due to the relaxation parameters. LM-MLDS provides a slightly higher $COV_{WM}$ at early iterations than the other methods but provides a lower $COV_{WM}$ at later iterations than LM-OSEM. **Fig. 4 top right** shows the $COV_{GM}$ curves in the simulation data with a full count. LM-MLDS provides a lower $COV_{GM}$ curve than the other methods. It suggests that the proximity term in Eq. (11) acts relatively strongly in the noisy hot region and relatively weakly in the less noisy cold region. **Fig. 4 bottom row** show the COV curves for the 1/20 count case. LM-OSEM provides higher COV curves than LM-

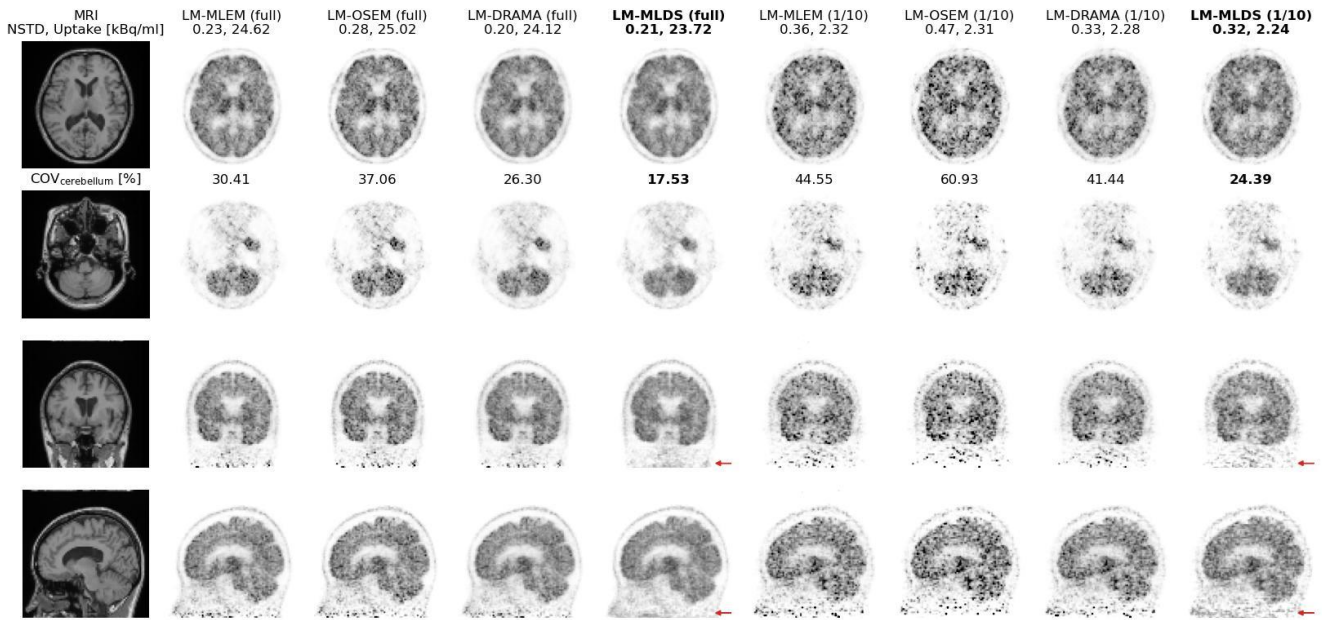

**Fig. 5.** Images of clinical data having full and 1/10 count, respectively, reconstructed using the proposed method and the other methods. From left to right, MRI, LM-MLEM (full), LM-OSEM (full), LM-DRAMA (full), LM-MLDS (full), LM-MLEM (1/10), LM-OSEM (1/10), LM-DRAMA (1/10), and LM-MLDS (1/10). Each image is tagged with NSTD, thalamus uptake, and $\text{COV}_{\text{cerebellum}}$. We show the images of LM-MLEM at 60 and 30 iterations, LM-OSEM, LM-DRAMA, and LM-DLS at 2 and 1 main iteration for full count and 1/10 count, respectively. The red arrow indicates the axial position where the false hot spot was suppressed by LM-MLDS.

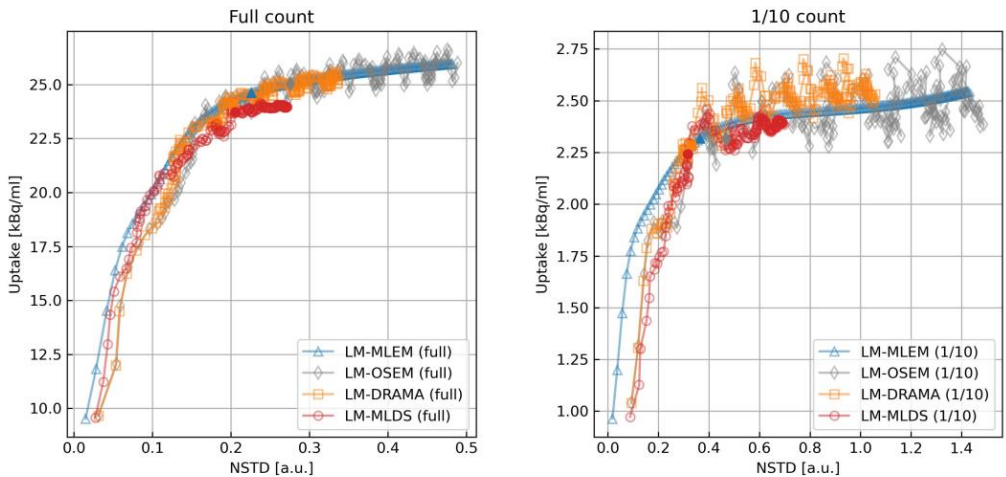

**Fig. 6.** Tradeoff curves between tumor uptake and NSTD on clinical data in (left) full and (right) 1/10 count, respectively. Fill markers correspond to the images shown in Fig. 5.

MLEM, probably because the low count expanded the effect of the limit cycle phenomenon.

**Fig. 5** shows the reconstructed images of the clinical data with full and 1/10 count, respectively. LM-MLDS provides less-noisy images in a transaxial slice, covering the whole brain from the top to the cerebellum, than the other methods (**Fig. 5, second row**). LM-MLDS achieves the lowest $\text{COV}_{\text{cerebellum}}$. When focusing on slices at the edge of the axial field of view (FOV) in the coronal and sagittal sections, it is evident that LM-MLDS suppresses false hot spots (**Fig. 5, red arrow**). This reason is discussed in the next section. **Fig. 6 left** shows the tradeoff curves between NSTD and thalamic uptake in clinical data with a full count. The LM-MLDS method provides a less noisy and rather more homogeneous appearance of $^{11}$C-MeQAA uptake in the thalamus than the other methods. This result suggests that LM-MLDS potentially mitigated the occurance of spotty high tracer accumulation in the α7 nAChR-rich thalamus in the clinical data. Indeed, the simulation data (**Figure 3, left**) further supports this result. Unlike the PSNR of the simulation data, LM-MLDS did not improve the NSTD relative to the other methods. Note that NSTD is evaluated in WM as a cold region, and because NSTD is evaluated based on the deviation of ROI values, it may be affected not only by noise but also by the accuracy of data correction. **Fig. 6 right** shows the tradeoff curves between NSTD and thalamus uptake in the clinical data with a 1/10 count. LM-MLDS reduces the oscillation of the tradeoff curve owing to the limit cycle phenomenon relative to LM-OSEM and LM-DRAMA.

## VI. Discussion

We propose list-mode maximum likelihood PET reconstruction using Dykstra-like splitting (LM-MLDS). LM-MLDS penalizes the distance between the reconstructed and reference images to converge the block iterative method without controlling the relaxation parameter. We evaluated LM-MLDS using simulations and clinical data from a brain PET scanner.

LM-MLDS provided a slightly higher $COV_{WM}$ in the early iterations (**Fig. 4, top left**) and remarkably lower $COV_{GM}$ curves than the other methods (**Fig. 4, right**). This is probably due to the noise properties of MLEM and Gaussian denoising induced by the proximity operator. The MLEM can be viewed as a gradient descent with a step size dependent on the local activity level. The convergence of MLEM is fast and slow in the hot and cold regions, respectively. This indicates that MLEM has high and low variances in the hot and cold regions, respectively [36]. The proximity operator in Eq. (13) can be viewed as a maximum a posteriori estimation:

$$C \cdot \exp\left(-L_q(D|x)\right) \exp\left(-\frac{1}{2\alpha}\left\|x - \left(x^{(k,q)} + y_q^{(k)}\right)\right\|^2\right), (27)$$

where $C$ is the normalization constant [37]. This can be interpreted as a denoising task of Gaussian noise with a mean of $x^{(k,q)} + y_q^{(k)}$. The strength of this denoising is inversely dependent on the step size $\alpha$ and independent of the local activity level. Hence, Gaussian denoising induced by the proximity operator may mitigate the activity dependence of the noise level in MLEM. Consequently, LM-MLDS exhibited different noise properties and a lower $COV_{GM}$ than the other methods. In addition, Gaussian denoising induced by the proximity operator may affect tumor regions with relatively high activity, suppressing overshoot in PSF modeling.

Coronal and sagittal slices of clinical images showed that LM-MLDS reduced false hotspots at the edge of the axial FOV (**Fig. 5, red arrow**). Similar to the simulation results, Gaussian denoising by the proximity operator likely suppressed false hotspots at the axial FOV edge caused by both the low geometric sensitivity and the leakage of random and scatter events from outside the axial FOV. This consideration was supported by an improvement in the image quality of the transaxial slices, including the cerebellum, using LM-MLDS (**Fig. 5, second row**). These results suggested the LM-MLDS may be useful for improving the robustness of geometric sensitivity variations in the 3D PET scanners, especially brain-dedicated PET [38], [39] and breast PET [40] scanners. In addition, improving the image quality of slices that include the cerebellum may aid kinetic analysis based on a simplified reference tissue model [41]. LM-MLDS provided less noisy and more homogenous uptake in the thalamus than the other methods in the clinical setting (**Fig. 6**). This suggests that LM-MLDS may have suppressed the overestimation of $^{11}$C-MeQAA uptake in the α7 nAChR-rich thalamus of the living human brain by suppressing noise in high-activity regions, comparable to the simulation results.

The amplitude of the oscillation of the tradeoff curves seemed to depend on the count per subset (**Figs. 3 and 6**). In the list-mode block iterative method, it may be necessary to adaptively control the relaxation parameters and step size, depending on the count per subset. Such adaptive control of the relaxation parameters and step size are potential future research topics. Proximity operators were originally used to optimize objective functions that include nondifferentiable terms, such as TV; however, in this study, we focused on the effectiveness of proximity operators for block optimization. In the near future, we will explore LM-MLDS by incorporating regularization terms such as nonlinear filter-based priors [42], [43] and deep-image priors [44]–[47]. Visualizing the noise properties of LM-MLDS using the sampling method of the list data [48] is also an interesting topic for future research.

LM-MLDS shares similarities with the list-mode stochastic primal-dual hybrid gradient descent (LM-SPDHG) [19], which is a memory-efficient algorithm for list-mode PET reconstruction with TV regularization. The main difference between LM-MLDS and LM-SPDHG is the space of the dual variables. LM-SPDHG sets the dual variable in the list-mode data space of $y \in \mathbb{R}^T$, whereas LM-MLDS sets it in the product space of images for each subset of $y \in \mathbb{R}^{M \times J}$ where $J$ is the number of voxels. When the number of events is larger than the number of dual variables of LM-MLDS, such as $T > M \times J$, the LM-MLDS algorithm is more efficient than LM-SPDHG. Furthermore, LM-MLDS is a simple and powerful solution for PET image reconstruction, effectively extending the traditional LM-OSEM through the integration of additional regularization.

## VII. Conclusion

We proposed list-mode maximum likelihood PET reconstruction using Dykstra-like splitting (LM-MLDS). LM-MLDS was evaluated using simulation and clinical data. LM-MLDS provided better tradeoffs between PSNR and tumor uptake ratio in a simulation study. LM-MLDS prevented false hotspots at the edge of the axial FOV in a clinical study. These results indicate that LM-MLDS can replace LM-OSEM and LM-DRAMA as a list-mode block iterative method for state-of-the-art PET scanners.

## Acknowledgment

The authors thank the members of the Fifth Research Group of the Central Research Laboratory, Hamamatsu Photonics K. K., for their kind support. This study was supported by JSPS KAKENHI (grant number: JP22K07762).

## References

[1] M. E. Phelps, *PET: Molecular Imaging and Its Biological Applications*. New York, NY, USA: Springer-Verlag, 2004. [Online]. Available: https://www.springer.com/la/book/9780387403595

[2] L. A. Shepp and Y. Vardi, "Maximum likelihood reconstruction for emission tomography," *IEEE Trans. Med. Imaging*, vol. 1, no. 2, pp. 113–122, 1982.

[3] H. M. Hudson and R. S. Larking, "Accelerated image reconstruction using ordered subsets of projection data," *IEEE Trans. Med. Imaging*, vol. 13, no. 4, pp. 601–609, 1994.

[4] J. Browne and A. B. de Pierro, "A row-action alternative to the EM algorithm for maximizing likelihood in emission tomography," *IEEE Trans. Med. Imaging*, vol. 15, no. 5, pp. 687-699, 1996.

[5] E. Tanaka and H. Kudo, "Subset-dependent relaxation in block iterative algorithms for image reconstruction in emission tomography," *Phys. Med. Biol.*, vol. 48, no. 10, 1405–1422, 2003.

[6] E. S. H. Neto and A. R. de Pierro, "Convergence results for scaled gradient algorithms in positron emission tomography," *Inverse Problems*, vol. 21, no. 6, pp. 1905–1914, 2005.

[7] D. W. Townsend, R. A. Isoradi, and B. Bendriem, "Volume imaging tomographs," in *The Theory and Practice of 3D PET*, B. Bendriem and D. W. Townsend, Ed. Dordrecht, The Netherlands, Kluwer Academic Publishers, 1998, pp. 111–132.

[8] B. A. Spencer et al., "Peformance evaluation of the uEXPLORER total-body PET/CT scanner based on NEMA NU 2-2018 woth additional tests to characterize PET scanners with a long axial field of view," *J. Nucl. Med.*, vol. 62, no. 6, pp. 861–870, 2021.

[9] A. J. Reader et al., "One-pass list-mode EM algorithm for high-resolution 3-D PET image reconstruction into large arrays," *IEEE Trans. Nucl. Sci.*, vol. 49, no. 3, pp. 693–699, 2002.

[10] A. Rahmim et al., "Statistical list-mode image reconstruction for the high resolution research tomograph," *Phys. Med. Biol.*, vol. 49, no. 18, pp. 4239–4258, 2004.

[11] P. L. Combettes and J.-C. Pesquet, "Proximal splitting methods in signal processing," in *Fixed-Point Algorithms for Inverse Problems in Science and Engineering. Springer Optimization and Its Applications*, vol. 49, H. Bauschke et al., Ed. New York, NY, Springer, pp. 185–212, 2011.

[12] L. I. Rudin, S. Oscher, and E. Fatemi, "Nonlinear total variation based noise removal algorithms," *Physica D*, vol. 6, no. 1–4, pp. 259–268, 1992.

[13] G. B. Passty, "Ergodic convergence to a zero of the sum of monotone operators in Hilbert space," *J. Math. Anal. Appl.*, vol. 72, no. 2, pp. 383–390, 1979.

[14] J. P. Boyle and R. L. Dykstra, "A method for finding projections onto the intersection of convex sets in Hilbert space," *Lecture Notes in Statistics*, vol. 37, pp. 28–47, 1986.

[15] S. P. Han, "A decomposition method and its application to convex programming," *Math. Oper. Res.*, vol. 14, no. 2, pp. 237–248, 1989.

[16] A. J. Reader et al., "Deep learning for PET image reconstruction," *IEEE Trans. Radiat. Plasma Med. Sci.*, vol. 5, no. 1, pp. 1-25, 2020.

[17] F. Hashimoto et al. "Deep learning-based PET image denoising and reconstruction: a review," *Radiol. Phys. Technol.*, vol. 17, pp. 24–46, 2024. [Online]. Available: https://doi.org/10.1007/s12194-024-00780-3

[18] M. J. Ehrhardt, P. Markiewiez, and C.-B. Schönlieb, "Faster PET reconstruction with non-smooth priors by randomization and preconditioning," *Phys. Med. Biol.*, vol. 64, no. 22, pp. 225019, 2019.

[19] G. Schramm and M. Holler, "Fast and memory-efficient reconstruction of sparse Poisson data in listmode with non-smooth priors with application of time-of-flight PET," *Phys. Med. Biol.*, vol. 67, no. 15, pp. 155020, 2022.

[20] H. Kim, K. Sadakata, and H. Kudo, "Unified framework to construct fast row-action-type iterative CT reconstruction methods with total variation using muti proximal splitting," in *Proceedings of the 6th international conference on biomedical signal and image processing*, Suzhou, China, pp. 65-71, 2021.

[21] K. Sadakata, H. Kim, and H. Kudo, "Unified approach to fast convergent row-action type iterative methods for PET image reconstruction using multi proximal splitting," *J. Image Graph.*, vol. 10, no. 2, pp. 82-87, 2022.

[22] M. Watanabe et al., "Performance evaluation of a high-resolution brain PET scanner using four-layer MPPC DOI detectors," *Phys. Med. Biol.*, vol. 62, no. 17, pp. 7148-7166, 2017.

[23] T. Nakayam and H. Kudo, "Derivation and implementation of ordered-subsets algorithms for list-mode PET data," *2005 IEEE Nucl. Sci. Symp. Conf. Rec.*, Fajardo, PR, USA, pp. 1950–1954, 2005.

[24] X. Cao, Q. Xie, and P Xiao, "A regularized relaxed ordered subset list-mode reconstruction algorithm and its preliminary application to undersampling PET imaging," *Phys. Med. Biol.*, vol. 60, no. 1, pp. 49-66, 2015.

[25] A. Rahmim et al., "Statistical dynamic image reconstruction in state-of-the-art high resolution PET," *Phys. Med. Biol.* vol. 50, no. 20, pp. 4887-4912, 2005.

[26] J. Qi, "Calculation of the sensitivity image in list-mode reconstruction for PET," *IEEE Trans. Nucl. Sci.*, vol. 53, no. 5, pp. 2746-2751, 2006.

[27] K. Lange, D. R. Hunter, and I. Yang, "Optimization transfer using surrogate objective functions," *J. Comput. Graph. Stat.*, vol. 9, no. 1, pp. 1-20, 2000.

[28] G. Wang and J. Qi, "Penalized likelihood PET image reconstruction using patch-based edge-preserving regularization," *IEEE Trans. Med. Imaging*, vol. 31, no. 12, pp. 2194-2204, 2012.

[29] Y. Berker and Y. Li, "Attenuation correction in emission tomography using the emission data—A review," *Med. Phys.*, vol. 43, no. 2, pp. 807-832, 2016.

[30] R. D. Badawi and P. K. Marsden, "Developments in component-based normalization for 3D PET," *Phys. Med. Biol.*, vol. 44, no. 2, pp. 571-594, 1999.

[31] C. C. Watson, "New, faster, image-based scatter correction for 3D PET," *IEEE Trans. Nucl. Sci.*, vol. 47, no. 4, pp. 1587-1594, 2000.

[32] D. Brasse et al., "Correction methods for random coincidences in fully 3D whole-body PET: Impact on data and image quality," *J. Nucl. Med.*, vol. 46, no. 5, pp. 859-867, 2005.

[33] D. L. Collins et al., "Design and construction of a realistic digital brain phantom," *IEEE Trans Med. Imaging*, vol. 17, no. 3, pp. 463–468, 1998.

[34] K. Nakaizumi et al., "In vivo depiction of α7 nicotinic receptor loss for cognitive decline in Alzheimer's disease," *J. Alzheimers Dis.*, vol. 61, no. 4, pp. 1355-1365.

[35] D. Kidera et al., "The edge artifact in the point-spread function-based PET reconstruction at different sphere-to-background ratios of radioactivity," *Ann. Nucl. Med.*, vol. 30, no. 2, pp. 97–103, 2016.

[36] D. W. Wilson, B. M. W. Tsui, and H. H. Barrett, "Noise properties of the EM algorithm. II. Monte Carlo simulations," Phys. Med. Biol., vol. 39, no. 5, pp. 847-871, 1994.

[37] K. Yatabe, "On the integration of proximal splitting algorithms and deep learning," *Bulletin of the Japan Society for Industrial and Applied Mathematics*, vol. 33, no. 1, pp. 14-24, 2023.

[38] H. Tashima and T. Yamaya, "Proposed helmet PET geometries with add-on detectors for high sensitivity brain imaging," *Phys. Med. Biol.*, vol. 61, no. 19, pp. 7205–7220, 2016.

[39] Y. Onishi et al., "Performance evaluation of dedicated brain PET scanner with motion correction system," *Ann. Nucl. Med.*, vol. 36, no. 8, pp. 746-755, 2022.

[40] D. Morimoto-Ishikawa et al., "Evaluation of performance of a high-resolution time-of-flight PET system dedicated to the head and breast according to NEMA NU 2-2012 standard," *EJNMMI Phys.*, vol. 9, no. 1, pp. 88, 2022. [Online]. Available. https://doi.org/10.1186/s40658-022-00518-3

[41] A. A. Lammerstma and S. P. Hume, "Simplified reference tissue model for PET receptor studies," *Neuroimage*, vol 4., no. 3, pp. 153–158, 1996.

[42] J. Dong and H. Kudo, "Proposal of compressed sensing using nonlinear sparsifying transform for CT image reconstruction," *Med. Imaging Tech.*, vol. 34, no. 5, pp. 235-243, 2016.

[43] Y. Romano, M. Elad, and P. Mianfar, "The little engine that could: Regularization by denoising (RED)," *SIAM J. Imaging Sci.*, vol. 10, no. 4, pp. 1804-1844, 2017.

[44] K. Gong et al., "PET image reconstruction using deep image prior," *IEEE Trans. Med. Imaging*, vol. 38, no. 7, pp. 1655-1665, 2018.

[45] F. Hashimoto, K. Ote, and Y. Onishi, "PET image reconstruction incorporating deep image prior and a forward projection model," *IEEE Trans. Radiat. Plasma Med. Sci.*, vol. 6, no. 8, pp. 841-846, 2022.

[46] S. Li et al., "Neural KEM: A kernel method with deep coefficient prior for PET image reconstruction," *IEEE Trans. Med. Imaging*, vol. 42, no. 3, pp. 785-796, 2023.

[47] K. Ote et al., "List-mode PET image reconstruction using deep image prior," *IEEE Trans. Med. Imaging*, vol. 42, no. 6, pp. 1822–1834, 2023.

[48] M. Dahlbom, "Estimation of image noise in PET using the bootstrap method," *IEEE Trans. Nucl. Sci.*, vol. 49, no. 5, pp. 2062–2066, 2002.